\documentclass[aps,pra,twocolumn,superscriptaddress,10pt,twoside,showpacs]{revtex4-2} 

\usepackage{amsmath,amssymb}
\usepackage{mathrsfs}
\usepackage{epstopdf}
\usepackage{graphicx}
\usepackage{bm,color,bbm}
\usepackage{natbib}
\usepackage{multirow}
\usepackage[hidelinks]{hyperref}
\usepackage{caption}
\usepackage[table, x11names]{xcolor}
\definecolor{aqua}{rgb}{0.7,1,1}
\definecolor{yellowish}{rgb}{1,1,0.7}
\definecolor{greenish}{rgb}{0.5,1,0.4}
\usepackage{subcaption}
\usepackage{fancyhdr}
\usepackage{multirow}
\usepackage{cancel}
\newcommand{\nn}{\nonumber}

\newcommand{\appropto}{\mathrel{\vcenter{\offinterlineskip\halign{\hfil$##$\cr\propto\cr\noalign{\kern2pt}\sim\cr\noalign{\kern-2pt}}}}}

\begin{document}
\fancyhead{} 
\fancyhead[LE,RO]{\ifnum\value{page}<2\relax\else\thepage\fi}

\title{Bypassing the filtering challenges in microwave-optical quantum transduction\\ through optomechanical four-wave mixing}

\author{James Schneeloch}
\email{james.schneeloch.1@afrl.af.mil}
\affiliation{Air Force Research Laboratory, Information Directorate, Rome, New York, 13441, USA}

\author{Erin Sheridan}
\affiliation{Air Force Research Laboratory, Information Directorate, Rome, New York, 13441, USA}

\author{A. Matthew Smith}
\affiliation{Air Force Research Laboratory, Information Directorate, Rome, New York, 13441, USA}

\author{Christopher C. Tison}
\affiliation{Air Force Research Laboratory, Information Directorate, Rome, New York, 13441, USA}

\author{Daniel L. Campbell}
\affiliation{Air Force Research Laboratory, Information Directorate, Rome, New York, 13441, USA}

\author{Matthew D. LaHaye}
\affiliation{Air Force Research Laboratory, Information Directorate, Rome, New York, 13441, USA}

\author{Michael L. Fanto}
\affiliation{Air Force Research Laboratory, Information Directorate, Rome, New York, 13441, USA}

\author{Paul M. Alsing}
\affiliation{Air Force Research Laboratory, Information Directorate, Rome, New York, 13441, USA}

\date{\today}

\begin{abstract}
Microwave-optical quantum transduction is a key enabling technology in quantum networking, but has been plagued by a formidable technical challenge. As most microwave-optical-transduction techniques rely on three-wave mixing processes, the processes consume photons from a driving telecom-band (pump) laser to convert input microwave photons into telecom-band photons detuned from the laser by this microwave frequency. However, cleanly separating out single photons detuned only a few GHz away from a classically bright laser in the same spatial mode requires frequency filters of unprecedented extinction over a very narrow transition band, straining the capabilities of today's technology. Instead of confronting this challenge directly, we show how one may achieve the same transduction objective with comparable efficiency using a four-wave mixing process in which \emph{pairs} of pump photons are consumed to produce transduced optical photons widely separated in frequency from the pump. We develop this process by considering higher-order analogues of photoelasticity and electrostriction than those used in conventional optomechanics, and examine how the efficiency of this process can be made to exceed conventional optomechanical couplings.
\end{abstract}


\maketitle
\thispagestyle{fancy}

\newpage

Quantum transduction between light and sound (converting between acoustic phonons and photons) is a critical intermediate process employed in the more promising technologies toward making large-scale quantum networking possible \cite{lauk2020perspectives, han2021microwave, awschalom2021development, xu2024optomechanical}. Of particular relevance to the subject of this paper, piezo-optomechanical transduction \cite{blesin2021quantum, weaver2024integrated, mirhosseini2020superconducting, han2020cavity} uses a piezoelectric interaction to convert microwave photons from superconducting-circuit (SC) qubits into acoustic phonons of the same frequency, and then an optomechanical interaction to convert these sound quanta into telecom-band photons by consuming pump photons from a laser to make up the energy difference. 

As a three-wave mixing process between two photons and one phonon, the optomechanical interaction these transducers employ introduces a formidable technical challenge in filtering. The difference in frequency between the transduction photons and the pump photons is about five orders of magnitude smaller than either one itself (e.g., a detuning of 2 GHz from a central frequency of 200 THz). Moreover, one must filter the pump aggressively enough that both no fringe pump photons are in the transduction band, and that the transduction band photons can be picked out without picking up residual pump photons. Accomplishing this directly requires developing filtering technology of unprecedented extinction for so narrow a transition band. Though highly dispersive filters exist that may cleanly separate spectral lines with similar detunings \cite{starling2012double, VIPAPatent, xiao2004dispersion}, and there are proposed schemes of concatenating multiple filters to achieve high extinction \cite{krastanov2021optically}, the simultaneous challenge of achieving the high extinction ratios needed for \emph{single-photon} quantum transduction is as yet unmet. In this Rapid Communication, we show how to circumvent this challenge and achieve comparable optomechanical couplings using a novel \emph{four}-wave mixing process between three photons and one acoustic phonon.

The coupling between sound (i.e., strain) in a material, and the electromagnetic energy passing through it has been studied in one form or another for more than 140 years \cite{voss2014march, osterberg1937piezodielectric, barocchi1971multiphoton, cleland2013foundations}. To first (linear) order in the mechanical and electric displacement fields, we have the forward and converse piezoelectric effects. To second order, we have electrostriction and photoelasticity, and it is at this order that we have the standard three-wave-mixing optomechanical interaction. At each order, these properties are connected thermodynamically by Maxwell relations, or generalizations thereof. 

Extrapolating to the third-order relationship between the mechanical and electric displacement fields, one may show (as is worked out in Appendix \ref{Appa}) that what we call the second-order photoelasticity (change in second-order nonlinear inverse permittivity proportional to the applied strain) is thermodynamically connected to a \emph{cubic} electrostriction (induced strain proportional to the cube of the applied electric field). Far from being an esoteric relation between physically insignificant factors, we demonstrate how this third-order connection can amount to a substantial optomechanical coupling without suffering from the same filtering challenges as in the standard case.

Although higher-order photon-phonon interactions were first examined over fifty years ago \cite{barocchi1971multiphoton}, the field remains largely unexplored, with the exception of using a second-order photoelasticity to examine stress-induced changes to second-order optical nonlinearity \cite{PhysRevB.62.13455} and in AMO-based ensembles \cite{zibrov2002four, covey2019microwave, borowka2024continuous, kumar2023quantum}. To our knowledge, our work herein represents the first study of this higher-order interaction in an optomechanical platform quantum transduction. 

\begin{figure}[t]
\centerline{\includegraphics[width=0.85\columnwidth]{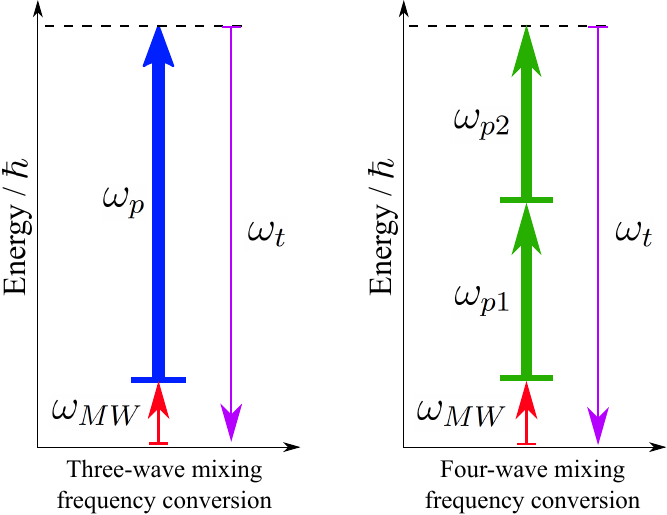}}
\caption{Energy-level diagram showing the difference difference between three- and four-wave mixing processes in frequency conversion from a GHz phonon at $\omega=\omega_{MW}$ to a telecom-band photon at $\omega=\omega_{t}$. The arrows in bold are classically bright fields at frequencies $\omega_{p}$, $\omega_{p1}$, and $\omega_{p2}$, respectively, which provide the required energy for the conversion to occur.}\label{Fconversion}
\end{figure}

In this Rapid Communication, we demonstrate the feasibility of microwave - telecom quantum transduction using optomechanical four-wave mixing. While the standard optomechanical three-wave mixing process consumes individual pump photons to facilitate transduction, optomechanical four-wave mixing consumes \emph{pairs} of pump photons to facilitate transduction. Where the transduction photons in this case have nearly twice the frequency of any one pump photon, both pump and transduced light can be easily filtered to high extinction with conventional optical filters.

Rather than give a full treatment of the entire transducer, we consider that the optomechanical coupling critical to the transduction process can be gauged by an effective photoelasticity. For the standard three-wave mixing-based coupling, we may use the photoelasticity as tabulated in tables of material parameters to quantify the strength of the effective coupling. For the case of our novel four-wave mixing, we may combine the second-order photoelasticity (defined presently) with one factor of the electric displacement field to obtain a comparable factor we call a \emph{virtual photoelasticity}. To facilitate this side-by-side comparison, we also develop a formula for the second-order photoelasticity using standard material parameters.

The optomechanical interaction in microwave-optical quantum transduction can be expressed by considering how the strain tensor $x_{ij}$ in a material alters its inverse permittivity $\eta_{ij}$. To first-order, we can describe this as:
\begin{equation}
\epsilon_{0}\eta_{ij}^{(\text{eff})}(x_{k\ell})\approx \epsilon_{0}\eta_{ij}(0) + p_{ijk\ell}^{(\text{eff})}x_{k\ell},
\end{equation}
where the effective photoelasticity tensor $p_{ijk\ell}^{(\text{eff})}$ is defined as \cite{AcoustoOpticsSalehTeich} the derivative of the \emph{relative} inverse permittivity $\epsilon_{0}\eta_{ij}$ with respect to strain $x_{k\ell}$ for small strain:
\begin{equation}\label{photoelast}
p_{ijk\ell}^{(\text{eff})}\equiv \epsilon_{0}\bigg(\frac{\partial \eta_{ij}^{(\text{eff})}}{\partial x_{k\ell}}\bigg|_{x\rightarrow 0}.
\end{equation} 
Note that the effective photoelasticity incorporates the true bulk photoelasticity of the material, as well as other contributions due to shifting boundaries, among other effects. Also, we will use the Einstein summation convention in this work for repeated indices (unless otherwise stated) to simplify notation. As part of the total hamiltonian for the electromagnetic field \cite{quesada2017you}, i.e.:
\begin{equation}
\hat{H}=\frac{1}{2}\int d^{3}r\left(\eta_{ij}\hat{D}_{i}\hat{D}_{j} +\frac{1}{\mu_{ij}}\hat{B}_{i}\hat{B}_{j}\right),
\end{equation}
(where the hats denote quantum operators), the strain-induced shift to $\eta_{ij}$ yields an additional term we call the optomechanical interaction hamiltonian:
\begin{equation}\label{hint1}
\hat{H}_{int}^{(1)}=\frac{1}{2\epsilon_{0}}\int d^{3}r\left( p_{ijk\ell}^{(\text{eff})}\hat{D}_{i}\hat{D}_{j}\hat{x}_{k\ell}\right),
\end{equation}
which is second-order in the electric displacement field $\hat{D}$, and first-order in the mechanical (acoustic) displacement $u_{i}$, by virtue of the strain $x_{ij}$ being given by the gradient of the displacement: $\frac{\partial \hat{u}_{i}}{\partial r_{j}}$, in the absence of rotational motion.

Just as the ordinary photoelasticity $p_{ijk\ell}$ describes the first-order change to $\eta_{ij}$ with respect to strain $x_{k\ell}$, we define the second-order photoelasticity $q_{hijk\ell}$ to be the second-order change to the relative permittivity $(\epsilon_{0}\eta_{hi})$ with respect to electric displacement field $D_{k}$ and strain $x_{k\ell}$:
\begin{equation}
q_{hijk\ell}\equiv \epsilon_{0}\bigg(\frac{\partial^{2}\eta_{hi}}{\partial x_{k\ell}\partial D_{j}}\bigg|_{(D,x)\rightarrow 0}.
\end{equation}
Note that while the ordinary photoelasticity $p_{ijk\ell}$ is dimensionless, the second-order photoelasticity $q_{ijk\ell m}$ has dimensions of $m^{2}/C$.

Between a vanishing electric field, and an electric field so large that it ionizes the solid, the inverse permittivity $\eta_{ij}$ is a nonlinear function of $D_{k}$ that can be represented by the following power series:
\begin{equation}\label{etaseries}
\eta_{ij}=\eta_{ij}^{(1)} + \eta_{ijk}^{(2)}D_{k} + \eta_{ijk\ell}^{(3)}D_{k}D_{\ell}+...,
\end{equation}
where for example the second-order inverse optical susceptibility is given by: 
\begin{equation}
\eta_{hij}^{(2)}\equiv \bigg(\frac{\partial\eta_{hi}}{\partial D_{j}}\bigg|_{D\rightarrow 0}.
\end{equation}
Here, we see that the second-order photoelasticity $q_{hijk\ell}$ is equal to the derivative of this second-order inverse susceptibility with respect to strain $x_{k\ell}$ (for small strain).

Following from Equations \eqref{hint1} and \eqref{etaseries}, the electromagnetic hamiltonian for a second-order nonlinear dielectric \cite{schneeloch2019introduction,quesada2017you} is given by:
\begin{align}
\hat{H}&=\frac{1}{2}\int d^{3}r\left(\eta_{ij}^{(1)}\hat{D}_{i}\hat{D}_{j} +\frac{1}{\mu_{ij}}\hat{B}_{i}\hat{B}_{j}\right)\\
&+\frac{1}{3}\int d^{3}r\left(\eta_{ijk}^{(2)}\hat{D}_{i}\hat{D}_{j}\hat{D}_{k}\right)\nn
\end{align}
If we include the effect of first and second-order photoelasticity, we obtain an additional higher-order optomechanical interaction term given by:
\begin{equation}\label{Hint2}
\hat{H}_{int}^{(2)}=\frac{1}{3\epsilon_{0}}\int d^{3}r\left( q_{hijk\ell}\hat{D}_{h}\hat{D}_{i}\hat{D}_{j}\hat{x}_{k\ell}\right)
\end{equation}
which is third-order in the electromagnetic field, and first order in the mechanical displacement field, making it a four-wave mixing interaction hamiltonian.

In order to understand the significance of this optomechanical four-wave mixing interaction, we must be able to estimate the value of $q_{hijk\ell}$ for different materials, and perform a side-by-side comparison of this with the standard three-wave mixing interaction \eqref{hint1}. To estimate the value of the second-order photoelasticity $q_{hijk\ell}$, we make use of Miller's rule in nonlinear optics \cite{miller1964optical,boyd2008nonlinear}, which approximates the second-order nonlinear susceptibility at the three frequencies being mixed as proportional to the product of the first-order susceptibilities at each respective frequency. In terms of inverse optical susceptibilities, Miller's rule takes the form: 
\begin{equation}\label{millersruleEta}
\eta^{(2)}_{hij}(\omega_{3}=\omega_{1}+\omega_{2})= -Q\prod_{n=1,2,3} \left(1-\epsilon_{0}\eta^{(1)}_{nn}(\omega_{n})\right)
\end{equation}
where $Q$ is a constant of proportionality (independent of frequency). If we assume the constant of proportionality $Q$ is a constant for each material (and that we are in a coordinate frame that diagonalizes $\eta_{ij}^{(1)}$), then we may obtain the approximate empirical formula for the second-order photoelasticity without needing to know the value of $Q$:
\begin{equation}
q_{hijk\ell}=\epsilon_{0}\frac{\partial \eta_{hij}^{(2)}}{\partial x_{k\ell}}\approx-\epsilon_{0}\eta^{(2)}_{hij}\sum_{n=1,2,3}\frac{p_{nnk\ell}(\omega_{n})}{1-\epsilon_{0}\eta_{nn}^{(1)}(\omega_{n})}
\end{equation}
This is accomplished by differentiating our expression for $\eta_{hij}^{(2)}$ (Eq.~\eqref{millersruleEta}) with respect to strain, and eliminating $Q$ in the resulting expression by substituting its value as determined by Eq.~\eqref{millersruleEta}. In terms of conventionally tabulated parameters, our expression for the second-order photoelasticity may be simplified and approximated to:
\begin{equation}
q^{(\text{eff})}\approx -\frac{2 d_{\text{eff}}}{\epsilon_{0}n^{2}(\omega_{1})n^{2}(\omega_{2})n^{2}(\omega_{3})}\!\!\sum_{n=1,2,3}\!\frac{p_{nnk\ell}(\omega_{n})}{1-\frac{1}{n^{2}(\omega_{n})}}    
\end{equation}
where the indices of refraction at the given frequencies are taken along the appropriate polarizations, and $d_{\text{eff}}$ is half the effective value of the second-order nonlinear optical susceptibility (accounting for quasi-phase matching factors as necessary \footnote{When accounting for quasi-phase matching where the medium is periodically poled to periodically flip the sign of the second-order nonlinear-optical susceptibility over the length of the medium, $d_{\text{eff}}$ decreases from its nominal value by a factor of $2/(n\pi)$ where $n$ is the diffraction order (ideally unity).}). Note that we use $d_{\text{eff}}$, instead of the inverse nonlinear-optical susceptibility because the latter is not commonly tabulated. For the transduction process under consideration, we let $(\omega_{1}=\omega_{p1},\omega_{2}=\omega_{p2}, \omega_{3}=\omega_{t}=\omega_{p1}+\omega_{p2}+\omega_{m})$.

With this formula, we can enter constants to see what a typical value for $q_{\text{eff}}$ might be. As an example, for Barium titanate, where ($n(2\pi c/1310nm)\approx 2.27$, $n(2\pi c/2600nm)\approx 2.26$ \cite{palik1997handbook}, $d_{\text{eff}}\approx 10 pm/V^{*}$ \cite{ZhouBTO3p1mumIEEE2023} \footnote{* This is an order of magnitude estimate based on SHG from 3.1-1.55 microns).} \footnote{Note: The electro-optic (Pockels) tensor and the $\chi^{(2)}$ nonlinear susceptibility tensor describe the same nonlinear-response, albeit in very different frequency bands. The Pockels tensor is tabulated at low (DC-GHz) frequencies, while the nonlinear susceptibilities is defined at much higher (optical) frequencies. The exceptionally large (of order $10^{3}pm/V$) electro-optic coefficient of barium titanate does not immediately imply a correspondingly large value for $d_{\text{eff}}$.}, $p_{1133}\approx 0.2^{**}$, $p_{2233}\approx 0.2^{**}$, $p_{3333}\approx 0.77^{**}$)\cite{zgonicPRB1994} \footnote{(** this is an order of magnitude estimate based on photoelasticity at 633nm)}, this gives an effective second-order photoelasticity $q^{(\text{eff})}\approx 2.45\times 10^{-2}m^{2}/C$. In what follows, we will show how the optomechanical coupling using second-order photoelasticity compares with conventional approaches.

Where the optomechanical three-wave mixing \eqref{hint1} and four-wave mixing interaction hamiltonians \eqref{Hint2} differ by only a few factors, we may group $q_{hijk\ell}$ with $\hat{D}_{h}$ and other constant factors to emulate a (virtual) first-order photoelasticity $p_{ijk\ell}^{(virt)}$:
\begin{equation}
\hat{H}_{int}^{(2)}=\frac{1}{2\epsilon_{0}}\int d^{3}r\left(p_{ijk\ell}^{(virt)}\hat{D}_{i}\hat{D}_{j}\hat{x}_{k\ell}\right)
\end{equation}
where
\begin{equation}
p_{ijk\ell}^{(virt)}=\frac{2}{3}q_{hijk\ell}^{(\text{eff})}\hat{D}_{h}=\frac{2}{3}\epsilon_{0}q_{hijk\ell}^{(\text{eff})}(\epsilon_{R})_{hm}\hat{E}_{m}
\end{equation}

For our example with barium titanate, we had a $q^{(\text{eff})}\approx 2.45\times 10^{-2}m^{2}/C$, and $\epsilon_{R}\approx 5.09$. In these units, $p_{ij\ell m}^{(virt)}\approx 7.35\times 10^{-13}\hat{E}_{i}$. Where the average electric field of a Gaussian beam of light propagating through a barium titanate waveguide with mode field diameter (MFD) of $1.2\mu m$ \footnote{The mode field diameter chosen here is an order-of magnitude estimate from step-index single-mode fiber optics assuming a Barium titanate core embedded in a silica cladding, where the core radius is chosen to minimize the MFD subject to a constant wavelength.} has an average peak electric field amplitude given by the relation:
\begin{equation}
|E|\approx\sqrt{\frac{16 P}{n\pi \epsilon_{0} c\, (\text{MFD})^{2}}}
\end{equation}
which within an order of magnitude at 1mW power would be approximately $7.68\times 10^{5}V/m$, corresponding to a peak intensity of $88.4 kW/cm^{2}$, well below its optical damage threshold of $0.54 GW/cm^2$ \cite{Mathey:2000} (about $6.11W$ for this beam diameter), \footnote{(for comparison, the breakdown field in air is only about $3\times10^{6}$V/m))}. As a general function of power $P$, these parameters give a virtual photoelasticity $p^{(virt)}$ of about $1.787\times 10^{-5}\sqrt{P}$. While this is rather small compared to the nominal photoelasticity of $0.77$ for barium titanate, it may still be sufficient for efficient quantum transduction to occur. For the transducer in \cite{blesin2021quantum}, the single-photon optomechanical coupling $g_{0}$ is about $2\pi\times400Hz$, where other optomechanical transducers may have couplings as high as $2\pi\times 850kHz$ \cite{Chiappina:23}, with the majority of this increase coming from design, rather than different bulk material properties. All other material parameters constant, and using the parameters of the transducer in \cite{blesin2021quantum} as a benchmark, then the rapid growth of the overall optomechanical coupling in optomechanical four-wave mixing relative to the standard approach shows that optimal transduction may be achieved with powers of the order $10^{1}W$. While this is currently above the damage threshold of Barium titanate, any further enhancements to the cumulative factor of nonlinearity and photoelasticity $(d_{\text{eff}}\;p_{nnk\ell})$ (e.g., by electrically polarizing the medium with a DC electric field \cite{PhysRevLett.8.404}), may enable near-unity transduction efficiencies at intensities below the optical damage threshold of the material.

Independent of bulk material properties, optomechanical four-wave mixing can only occur if phase-matching conditions similar to nonlinear-optical four-wave mixing are satisfied. Assuming an approximately rectangular geometry with a periodic poling to compensate for any momentum offset, the effective optomechanical coupling due to second-order photoelasticity is proportional to the phase-matching integral:
\begin{equation}
g_{\text{eff}}\propto \left(\int dz e^{i\Delta k_{z} z}\right)
\end{equation}
where $z$ ranges from zero to $L$, the length of the medium, and:
\begin{align}
 \Delta k_{z} &= k_{tz}-k_{p1z}-k_{p2z}-k_{mz}-k_{\Lambda z}\\
 &\approx\frac{n(\omega_{t})\omega_{t}}{c}-\frac{n(\omega_{p1})\omega_{p1}}{c}-\frac{n(\omega_{p2})\omega_{t2}}{c}-\frac{\omega_{m}}{v_{sm}}-\frac{2\pi}{\Lambda}\nn
\end{align}
where $v_{sm}$ is the speed of sound in the material for this acoustic mode. As a side benefit of this process, the conditions achieving the correct phase-matching in optomechanical four-wave mixing, will generally not also satisfy the phase-matching conditions for optomechanical three-wave mixing. Because of this, the lower-order optomechanical coupling used in standard transducers will be suppressed at the same time that optomechanical four-wave mixing is optimized, removing it as a source of loss/noise.

In this work, we have shown how the frequency filtering challenges in quantum transduction can be circumvented by utilizing optomechanical four wave mixing. Moreover, we have given arguments for why this higher-order process ought to amount to comparable optomechanical coupling to the standard case. The benefits of this approach are not without their drawbacks, however. While optomechanical four-wave mixing has a power-scaling advantage in that its efficiency literally grows with pump power, its efficiency also varies with the phase of the pump, suggesting that additional stabilization may be required. Moreover, there will be additional optical design requirements for the transducer to be able to host a pair of wavelengths nearly an octave apart with maximal overlap/coupling. Independent of quantum transduction, this higher-order optomechanical connection may enable sensing technologies where using standard Brillouin scattering is impractical.

\begin{acknowledgments}
We gratefully acknowledge support from our colleagues at the Air Force Research Laboratory, and at Purdue University.

The views expressed are those of the authors and do not reflect the official guidance or position of the United States Government, the Department of Defense or of the United States Air Force. The appearance of external hyperlinks does not constitute endorsement by the United States Department of Defense (DoD) of the linked websites, or of the information, products, or services contained therein. The DoD does not exercise any editorial, security, or other control over the information you may find at these locations.
\end{acknowledgments}

\bibliography{EPRbib16}

\appendix
\begin{widetext}
\newpage
\section{Thermodynamic relations between electric and mechanical displacement}\label{Appa}
Consider a dielectric solid at constant temperature $T$. With heat free to flow in/out of this system, and wanting temperature as an independent variable, we employ the Helmholtz free energy $A=U-TS$ instead of the internal energy $U$ to describe the thermodynamics of this system. The differential change in its Helmholtz free energy $A$ is given by: 
\begin{equation}
dA=-SdT+X_{ij}dx_{ij}+E_{i}dD_{i},
\end{equation}
where $S$ is the entropy, $X_{ij}$ is the mechanical stress tensor, $x_{ij}$ is the mechanical strain tensor, $E_{i}$ is the ordinary electric field, and $D_{i}$ is the electric displacement field. In addition, $-SdT$ is the differential heat flow into/out of the system; $X_{ij}dx_{ij}$ is the differential mechanical work done on/by the system; and $E_{i}dD_{i}$ is the differential electromagnetic work done on/by the system, where electromagnetic energy can be stored in a dielectric by induced polarization. Just as in the body of our work, we use the Einstein summation convention on repeated indices to condense notation, where for example $E_{i}dD_{i}$ is an inner product between the electric field, and the differential change in the electric displacement field.

With temperature $T$ being constant, and because we can express $A$ as a function of $(T,x_{ij},D_{i})$, we may in turn express the mechanical stress tensor $X_{ij}$ as a function of strain $x_{ij}$, and the electric displacement field $D_{i}$:
\begin{equation}
X_{ij}=\left(\frac{\partial A}{\partial x_{ij}}\right)_{(T,\bar{x},D)}=X_{ij}(x,D).
\end{equation}
Similarly, we may express the Electric field $E_{i}$ as a function of these variables as well:
\begin{equation}
E_{i}=\left(\frac{\partial A}{\partial D_{i}}\right)_{(T,\bar{x},D)}=E_{i}(x,D).
\end{equation}
Both $X_{ij}$ and $E_{i}$ can be Taylor-expanded in powers of $x_{ij}$ and $D_{i}$. Since even relatively intense sound results in small displacements, we write out the Taylor expansion up to first order in strain $x_{ij}$ and up to third order in the electric displacement field $D_{i}$ (accounting for orders of strain):
\begin{align}
X_{ij}(x,D)&\approx\!\!
\begin{bmatrix}
const & + & \left(\frac{\partial X_{ij}}{\partial D_{k}}\right|_{0,0}\!\!\!\!\!D_{k} & + &\frac{1}{2}\left(\frac{\partial^{2} X_{ij}}{\partial D_{k}\partial D_{\ell}}\right|_{0,0}\!\!\!\!\!D_{k}D_{\ell} & + & \frac{1}{3!}\left(\frac{\partial^{3} X_{ij}}{\partial D_{k}\partial D_{\ell}\partial D_{m}}\right|_{0,0}\!\!\!\!\!D_{k}D_{\ell}D_{m}\\
+\left(\frac{\partial X_{ij}}{\partial x_{k\ell}}\right|_{0,0}\!\!\!\!\!x_{k\ell} & + & \frac{1}{2}\left(\frac{\partial^{2}X_{ij}}{\partial x_{k\ell}\partial D_{m}}\right|_{0,0}\!\!\!\!\!x_{k\ell}D_{m} & + & \frac{1}{3!}\left(\frac{\partial^{3} X_{ij}}{\partial x_{k\ell}\partial D_{m}\partial D_{n}}\right|_{0,0}\!\!\!\!\!x_{k\ell}D_{m}D_{n} & + & ... 
\end{bmatrix}\\
E_{i}(x,D)&\approx\!\!
\begin{bmatrix}
const & + & \left(\frac{\partial E_{i}}{\partial D_{k}}\right|_{0,0}\!\!\!\!\!D_{k} & + &\frac{1}{2}\left(\frac{\partial^{2} E_{i}}{\partial D_{k}\partial D_{\ell}}\right|_{0,0}\!\!\!\!\!D_{k}D_{\ell} & + & \frac{1}{3!}\left(\frac{\partial^{3} E_{i}}{\partial D_{k}\partial D_{\ell}\partial D_{m}}\right|_{0,0}\!\!\!\!\!D_{k}D_{\ell}D_{m}\\
+\left(\frac{\partial E_{i}}{\partial x_{k\ell}}\right|_{0,0}\!\!\!\!\!x_{k\ell} & + & \frac{1}{2}\left(\frac{\partial^{2}E_{i}}{\partial x_{k\ell}\partial D_{m}}\right|_{0,0}\!\!\!\!\!x_{k\ell}D_{m} & + & \frac{1}{3!}\left(\frac{\partial^{3} E_{i}}{\partial x_{k\ell}\partial D_{m}\partial D_{n}}\right|_{0,0}\!\!\!\!\!x_{k\ell}D_{m}D_{n} & + & ... 
\end{bmatrix}
\end{align}

In terms of conventional tensors (where we truncate the power series so terms equivalent to second and higher order derivatives of the electric field with respect to strain are neglected), these expressions can be simplified as:

\begin{align}
X_{ij}(x,D)&\approx\!\!
\begin{bmatrix}
const & + & (h_{ijk})D_{k} & + &\frac{1}{2}\left(\frac{p_{ijk\ell}}{\epsilon_{0}}\right)D_{k}D_{\ell} & + & \frac{1}{3!}\left(\frac{q_{ijk\ell m}}{\epsilon_{0}}\right)D_{k}D_{\ell}D_{m}\\
+\left(c_{ijk\ell}\right)x_{k\ell} & + & ... 
\end{bmatrix}\\
E_{i}(x,D)&\approx\!\!
\begin{bmatrix}
const & + & \eta_{ik}^{(1)}D_{k} & + & \eta_{ik\ell}^{(2)} D_{k}D_{\ell} & + & \eta_{ik\ell m}^{(3)}D_{k}D_{\ell}D_{m}\\
+(h_{k\ell i})x_{k\ell} & + & \frac{1}{2}\left(\frac{p_{mik\ell}}{\epsilon_{0}}\right)x_{k\ell}D_{m} & + & \frac{1}{3!}\left(\frac{2q_{k\ell mni}}{\epsilon_{0}}\right)x_{k\ell}D_{m}D_{n} & + & ... 
\end{bmatrix}
\end{align}
Here, $h_{ijk}$ is the stress-voltage form of the piezoelectric tensor \footnote{Note that the more common stress-charge form of the piezoelectric tensor is $e_{ijk}$ such that $h_{ijk}=\eta_{im}e_{mjk}$.}, $p_{ijk\ell}$ is the photoelasticity tensor, and $q_{ijk\ell m}$ is what we are defining as the second-order photoelasticity or cubic electrostriction tensor.

The relations between different orders of electric and mechanical displacement are based on the equality of different orders of mixed partial derivatives of $A$ (as define the Maxwell relations).

Where stress and the electric field are different first-derivatives of $A$, the first-order connection between mechanical stress and electric displacement is based on mixing second derivatives of $A$:
\begin{subequations}
\begin{align}
(h_{ijk})&\equiv \left(\frac{\partial X_{ij}}{\partial D_{k}}\right)\\
h_{ijk}&=\left(\frac{\partial^{2}A}{\partial D_{k}\partial x_{ij}}\right)=\left(\frac{\partial^{2}A}{\partial x_{ij}\partial D_{k}}\right)=\left(\frac{\partial E_{k}}{\partial x_{ij}}\right)
\end{align}
\end{subequations}
Indeed, this relation defines the connection between forward and converse piezoelectricity.

Beyond simple piezoelectricity, we have that the second-order connection between mechanical stress and electric displacement is based on mixing third derivatives of $A$:
\begin{subequations}
\begin{align}
\left(\frac{\partial^{2}X_{ij}}{\partial D_{k}\partial D_{\ell}}\right)&=\left(\frac{\partial^{3}A}{\partial D_{k}\partial D_{\ell}\partial x_{ij}}\right)=\left(\frac{\partial^{3}A}{\partial x_{ij}\partial D_{k}\partial D_{\ell}}\right)\nn\\
&=\left(\frac{\partial^{2} E_{\ell}}{\partial x_{ij}\partial D_{k}}\right)=\left(\frac{\partial\eta_{\ell k}}{\partial x_{ij}}\right)\equiv\left(\frac{p_{k\ell ij}}{\epsilon_{0}}\right)
\end{align}
\end{subequations}
Where $p_{ijk\ell}$ is the photoelasticity tensor, this second order-connection illustrates that it is related to a dependence of generated stress as a quadratic function of the electric field (i.e., electrostriction). This mechanism defines the conventional optomechanical coupling in transducers.

Beyond both piezoelectricity and standard electrostriction, we find third-order connection between mechanical stress and electric displacement based on mixing fourth derivatives of $A$:
\begin{subequations}
\begin{align}
\left(\frac{\partial^{3} X_{ij}}{\partial D_{k}\partial D_{\ell}\partial  D_{m}}\right)&=\left(\frac{\partial^{4}A}{\partial D_{k}\partial D_{\ell}D_{m}\partial x_{ij}}\right)=\left(\frac{\partial^{4}A}{\partial x_{ij}\partial D_{k}\partial D_{\ell}\partial D_{m}}\right)\nn\\
&=\left(\frac{\partial^{3} E_{m}}{\partial x_{ij}\partial D_{k}\partial  D_{\ell}}\right)=2\left(\frac{\partial\eta^{(2)}_{m \ell k}}{\partial x_{ij}}\right)\equiv 2\frac{q_{ijk\ell m}}{\epsilon_{0}}
\end{align}
\end{subequations}
Note: The factor of $2$ here comes from having to apply the product rule in the power series for $E_{i}=\eta_{ij}D_{j}$ (see equation \eqref{etaseries}  for $\eta_{ij}$ series) when differentiating:
\begin{equation}
\frac{\partial^{2}E_{m}}{\partial D_{\ell}\partial D_{k}}=\eta_{m\ell k}^{(2)}+\eta_{mk\ell}^{(2)}
\end{equation}
and assuming $\eta_{m\ell k}^{(2)}=\eta_{mk\ell}^{(2)}$ (such as the case of Kleinman symmetry where a lossless medium's nonlinear susceptibility is approximately independent of frequency \cite{boyd2008nonlinear}), and that the third-order susceptibility is negligible by comparison. This illustrates that the cubic electrostriction $q_{ijk\ell m}$ is up to a factor of two, equal to the second-order photoelasticity.

\end{widetext}
\end{document}